\documentclass[
  journal=jacsat,
  manuscript=article,
    layout=sincolumn
]{achemso}

\usepackage[version=3]{mhchem} 

\usepackage{booktabs}
\usepackage{siunitx}

\usepackage{titlesec}

\titleformat{\section}
  {\large\bfseries}
  {}      
  {0pt}
  {}

\titlespacing*{\section}
  {0pt}
  {1.2ex} 
  {0.8ex} 

\titleformat{\subsection}
  {\normalsize\bfseries} 
  {}      
  {0pt}
  {}

\titlespacing*{\subsection}
  {0pt}
  {1.0ex}
  {0.6ex}

\titleformat{\paragraph}
  {\footnotesize\bfseries}
  {}      
  {0pt}
  {}

\titlespacing*{\paragraph}
  {0pt}
  {0.6ex}
  {0.3ex}

\title{Hypothesis-and-Refinement Learning of Organic Structures from Multimodal Spectroscopic Data}

\author{Chengchun Liu}
\altaffiliation{These authors contributed equally.}
\affiliation{\textit{School of AI for Science, Peking University Shenzhen Graduate School, Shenzhen 518055, China}}
\alsoaffiliation{State Key Laboratory of Advanced Waterproof Materials, School of Materials Science and Engineering, Peking University, Beijing 100871, China}

\author{Zhiyuan Yan}
\altaffiliation{These authors contributed equally.}
\affiliation{\textit{School of AI for Science, Peking University Shenzhen Graduate School, Shenzhen 518055, China}}
\alsoaffiliation{School of Electronic and Computer Engineering, Peking University Shenzhen Graduate School, Shenzhen 518055, China}

\author{Li Yuan}
\email{yuanli-ece@pku.edu.cn}
\affiliation{\textit{School of AI for Science, Peking University Shenzhen Graduate School, Shenzhen 518055, China}}
\alsoaffiliation{School of Electronic and Computer Engineering, Peking University Shenzhen Graduate School, Shenzhen 518055, China}
\alsoaffiliation{Peng Cheng Laboratory, Shenzhen 518000, China}

\author{Hao Li}
\affiliation{\textit{School of AI for Science, Peking University Shenzhen Graduate School, Shenzhen 518055, China}}
\alsoaffiliation{School of Electronic and Computer Engineering, Peking University Shenzhen Graduate School, Shenzhen 518055, China}

\author{Boxuan Zhao}
\affiliation{\textit{School of AI for Science, Peking University Shenzhen Graduate School, Shenzhen 518055, China}}
\alsoaffiliation{State Key Laboratory of Advanced Waterproof Materials, School of Materials Science and Engineering, Peking University, Beijing 100871, China}

\author{Yonghong Tian}
\affiliation{\textit{School of AI for Science, Peking University Shenzhen Graduate School, Shenzhen 518055, China}}
\alsoaffiliation{School of Electronic and Computer Engineering, Peking University Shenzhen Graduate School, Shenzhen 518055, China}
\alsoaffiliation{Peng Cheng Laboratory, Shenzhen 518000, China}

\author{Bartosz A. Grzybowski}
\email{nanogrzybowski@gmail.com}
\affiliation{\textit{Institute of Organic Chemistry, Polish Academy of Sciences, Warsaw 01-224, Poland}}
\alsoaffiliation{IBS Center for Algorithmic and Robotized Synthesis (CARS), UNIST, Ulsan 44919, South Korea}
\alsoaffiliation{Department of Chemistry, UNIST, Ulsan 44919, South Korea}

\author{Fanyang Mo}
\email{fmo@pku.edu.cn}
\affiliation{\textit{School of AI for Science, Peking University Shenzhen Graduate School, Shenzhen 518055, China}}
\alsoaffiliation{State Key Laboratory of Advanced Waterproof Materials, School of Materials Science and Engineering, Peking University, Beijing 100871, China}
\alsoaffiliation{School of Advanced Materials, Peking University Shenzhen Graduate School, Shenzhen 518055, China}
\alsoaffiliation{Guangdong Provincial Key Laboratory of Nano-Micro Materials Research, Peking University Shenzhen Graduate School, Shenzhen 518055, China}

\begin{document}

\maketitle

\begin{abstract}
Determining molecular structures from spectroscopic data remains fundamentally challenging because the inverse problem is intrinsically underdetermined: individual spectra are sparse, low-dimensional, and encode only partial structural evidence relative to the vast space of possible molecules. We address this challenge by formulating automated structure elucidation as a scalable hypothesis--refinement paradigm that tightly integrates spectral evidence with large-scale molecular priors. To supply structure-resolving NMR signals for multimodal learning, we construct \textbf{QM9SPIN}, a DFT-derived dataset comprising diverse 1D and 2D spectra, including J-coupling, DEPT experiments, and explicit spin--spin interactions. On this foundation, we introduce \textbf{SpectroMol}, a spectrum-to-structure model that proposes chemically valid molecular hypotheses conditioned on multimodal spectral inputs. Complementarily, we develop \textbf{MS-Mol2Mol}, a high-resolution mass-constrained molecular generator that integrates molecular formula, exact mass, and degree of unsaturation within a conditional generative prior trained on 400 million molecules, ensuring global compositional consistency and chemically realistic refinement. The integrated system achieves 93.8\% top-1 accuracy on the simulated benchmark, adapts effectively from simulated to experimental spectra with limited experimental fine-tuning, and further improves experimental predictions through mass-guided refinement, establishing a scalable route toward automated, data-driven organic structure elucidation.
\end{abstract}

\section*{Keywords}
multimodal spectroscopy; structure elucidation; molecular generation; mass spectrometry; NMR

\section{Introduction}
\label{introduction}

\begin{figure}[t!]
    \centering
    \includegraphics[width=0.86\linewidth]{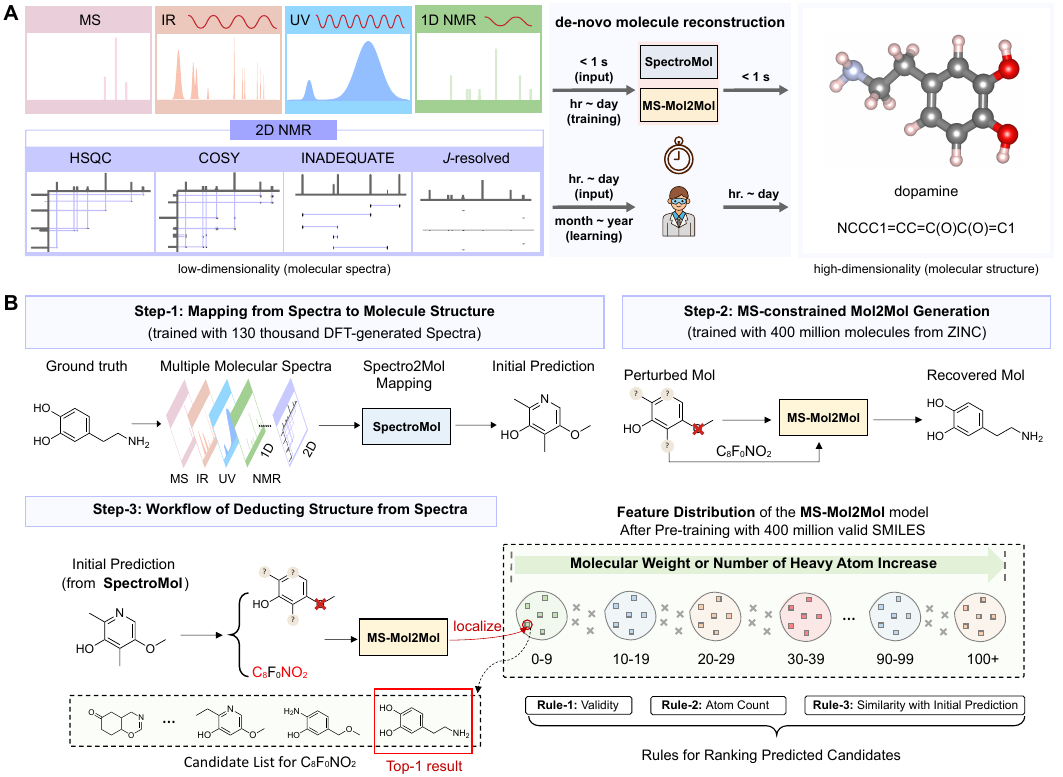}
    \caption{Overview of the multimodal structure-elucidation problem and workflow. (\textbf{A}) Molecular structure reconstruction from integrated spectroscopic evidence, including MS, IR, UV, and 1D/2D NMR, remains a demanding inverse problem because each modality contributes only partial structural constraints. (\textbf{B}) Example workflow using dopamine: SpectroMol first proposes structure hypotheses from multimodal spectra, after which MS-Mol2Mol refines these hypotheses under high-resolution mass constraints and the resulting candidates are filtered and ranked. }
    \label{fig_1}
\end{figure}

Determining molecular structures from spectroscopic data lies at the heart of chemistry, underpinning advances in synthetic methodology, drug discovery, and materials design~\cite{nicolaou2005chasing,barone2021computational}. Spectroscopic techniques---including one- and two-dimensional nuclear magnetic resonance (NMR), mass spectrometry (MS), infrared (IR), and ultraviolet--visible (UV--Vis) spectroscopy---encode diverse yet complementary information about molecular composition and bonding environments (Fig.~\ref{fig_1}A). Traditionally, chemists integrate fragmentary spectral cues through iterative reasoning to reconstruct a molecular framework. While effective, this manual process is time-consuming, subjective, and increasingly incompatible with high-throughput experimentation~\cite{liu2017unequivocal,felli202213c,liu2025automation} and autonomous synthesis platforms~\cite{dai2024autonomous,blair2022automated,chatterjee2020automated}. Accelerating and formalizing this reasoning process is therefore essential for realizing data-driven and autonomous chemical discovery.

The relationship between molecules and their spectra can be viewed as a bidirectional mapping. Considerable progress has been made in forward prediction---inferring spectra from known structures---through quantum-chemical simulation and machine learning~\cite{zou2023deep,li2024decoupled,long2025computed,liu2024infrared,han2022scalable,lou2026unsupervised,shi2026infrared,liu2026cross}. In contrast, the inverse problem (spectrum-to-structure) remains fundamentally challenging. Spectroscopic measurements are sparse, noisy, and modality-specific, whereas the chemical space of possible structures is vast and combinatorial~\cite{reif2021solid,laws2002solid,kalsi2007spectroscopy,bifulco2007determination,smith2018infrared,watson2007introduction}. Each modality provides only partial constraints---MS informs elemental composition but not connectivity; NMR encodes local chemical environments without uniquely fixing global topology; IR or UV--Vis identify functional motifs without structural context. Fundamentally, the inverse problem is not merely a data fusion task, but a constrained search over chemical space under incomplete and heterogeneous evidence.

Recent years have witnessed rapid advances in machine learning--based automated structure elucidation. For NMR spectroscopy, reinforcement learning and tree-search strategies iteratively construct candidate molecular graphs guided by predicted spectral features~\cite{sridharan2022deep}. Transformer-based architectures trained on synthetic spectra have demonstrated direct translation from $^1$H and $^{13}$C spectra to molecular representations~\cite{alberts2023learning}, and multitask models integrating convolutional and sequence components have enabled efficient reconstruction from routine one-dimensional NMR data without prior knowledge of molecular formula~\cite{hu2024accurate}. Automated systems such as DP4-AI streamline peak assignment and stereochemical validation from raw spectra~\cite{howarth2020dp4}, while embedding-based frameworks combining multimodal representations with discrete optimization align spectral evidence and molecular candidates within learned latent spaces~\cite{mirza2024spec2struct}. Parallel developments have emerged for infrared spectroscopy. Convolutional neural networks have been employed to identify functional groups directly from FTIR spectra~\cite{enders2021functional,jung2023automatic}, and contrastive learning frameworks have aligned IR embeddings with molecular representations to enable library ranking and generative structure prediction~\cite{kanakala2024spectra}. Collectively, these approaches demonstrate that data-driven models can extract meaningful structural information from individual spectroscopic modalities. Related efforts in mass spectrometry and metabolomics have likewise shown that database-guided fingerprint inference and unsupervised substructure discovery can greatly expand structure annotation from fragmentation data~\cite{duhrkop2015searching,vanderhooft2016topic}. These advances highlight the value of chemical priors, but they generally do not reconstruct complete molecular connectivity from heterogeneous spectra within a unified generative framework.

Despite this progress, most existing methods formulate spectrum-to-structure prediction as a single-stage mapping problem, translating spectral inputs directly into molecular representations. While effective within constrained domains, such formulations implicitly assume that all relevant evidence is simultaneously available and that structural reconstruction can be completed in a single inference step. In practice, however, spectroscopic reasoning is inherently iterative: provisional hypotheses are formed from partial evidence and progressively refined as additional constraints are incorporated. Moreover, end-to-end models trained solely on limited spectral datasets must implicitly learn both evidence interpretation and chemical plausibility within the same parameter space, making it difficult to navigate the vast combinatorial chemical space under heterogeneous constraints. These considerations suggest that effective multimodal structure elucidation may require reformulating the inverse problem itself---explicitly decoupling hypothesis generation from large-scale chemical prior modeling and framing structure determination as constrained generative reasoning over chemical space.

We therefore reformulate multimodal structure elucidation as a hypothesis--refinement problem. To enable this formulation, we construct \textbf{QM9-SPIN}, a DFT-augmented multimodal benchmark that extends QM9 molecules~\cite{ramakrishnan2014quantum} with systematically simulated and statistically characterized spectral modalities. Infrared and ultraviolet--visible spectra are adopted from the high-quality QM9s resource of Zou \emph{et al.}~\cite{zou2023deep}, while mass spectrometry and both one- and two-dimensional NMR signals are computed within a unified quantum-chemical pipeline. These modalities---spanning IR, UV--Vis, MS (Supporting Information, Figs.~S1--S3), molecular-level compositional statistics derived from SMILES (Supporting Information, Fig.~S4), and 1D/2D NMR experiments including $^{13}$C-NMR with DEPT annotation, $^{1}$H-NMR, COSY, HSQC, J-resolved spectroscopy, and INADEQUATE (Supporting Information, Figs.~S5--S11)---transform sparse spectral observables into structured local and relational constraints over molecular graph space.

Building on this foundation, we develop a unified two-stage framework inspired by human spectroscopic reasoning (Fig.~\ref{fig_1}B). In the first stage, a SpectroMol transformer integrates heterogeneous spectral evidence to generate candidate molecular hypotheses by learning cross-modal dependencies. In the second stage, MS-Mol2Mol refines these hypotheses under explicit compositional constraints derived from high-resolution mass spectrometry---including molecular formula, exact mass, and degree of unsaturation---while leveraging a conditional molecular generator pretrained on 400 million structures from ZINC20~\cite{irwin2020zinc20}. As illustrated in Supporting Information, Fig.~S12, this refinement stage concentrates the search within a composition-consistent region of chemical space while preserving spectrally induced structural motifs, thereby enabling formula-aware generative reasoning.

Our framework achieves 93.8\% top-1 accuracy on the QM9 benchmark and extends to experimental spectra through data-efficient adaptation and mass-guided refinement. Rather than treating spectral interpretation as a direct translation problem, our approach models structure elucidation as constrained generative reasoning guided by heterogeneous experimental evidence and large-scale molecular priors. This formulation establishes a computational paradigm for multimodal inverse chemical problems, illustrating how deep learning and quantum chemistry can jointly transform empirical spectral analysis into a systematic, predictive, and scalable process for automated molecular discovery.

\begin{table}[t!]
\centering
\caption{Performance comparison under different spectral input conditions. Metric definitions are provided in Supporting Information, section S4.}
\label{tab:spectrum_comparison}
\setlength{\tabcolsep}{4.5pt}
\renewcommand{\arraystretch}{1.15}
\footnotesize
\begin{tabular}{lccccccccc}
\toprule
Spectrum condition &
BLEU &
\shortstack{MACCS\\sim.} &
\shortstack{RDKit\\sim.} &
\shortstack{Morgan\\sim.} &
\shortstack{MCS\\ratio} &
\shortstack{MCS\\Tani.} &
\shortstack{MCS\\coeff.} &
\shortstack{Levenshtein} &
\shortstack{Top-1\\acc.} \\
\midrule
IR & 0.264 & 0.884 & 0.754 & 0.687 & 0.901 & 0.854 & 0.901 & 2.580 & 1.22\% \\
IR + UV & 0.280 & 0.263 & 0.124 & 0.116 & 0.549 & 0.403 & 0.554 & 8.341 & 1.28\% \\
IR + UV + MS & 0.368 & 0.468 & 0.236 & 0.173 & 0.621 & 0.486 & 0.621 & 7.163 & 13.4\% \\
\addlinespace[1pt]
\textbf{IR + UV + MS + NMR} & \textbf{0.970} & \textbf{0.994} & \textbf{0.985} & \textbf{0.981} & \textbf{0.995} & \textbf{0.993} & \textbf{0.995} & \textbf{0.438} & \textbf{93.8\%} \\
\bottomrule
\end{tabular}
\end{table}

\section{Results and Discussion}
\label{results_discussion}

\subsection{Propose--then--refine for multimodal structure elucidation}
\label{methods_framework}

\begin{figure}[t!]
    \centering
    \includegraphics[width=0.86\linewidth]{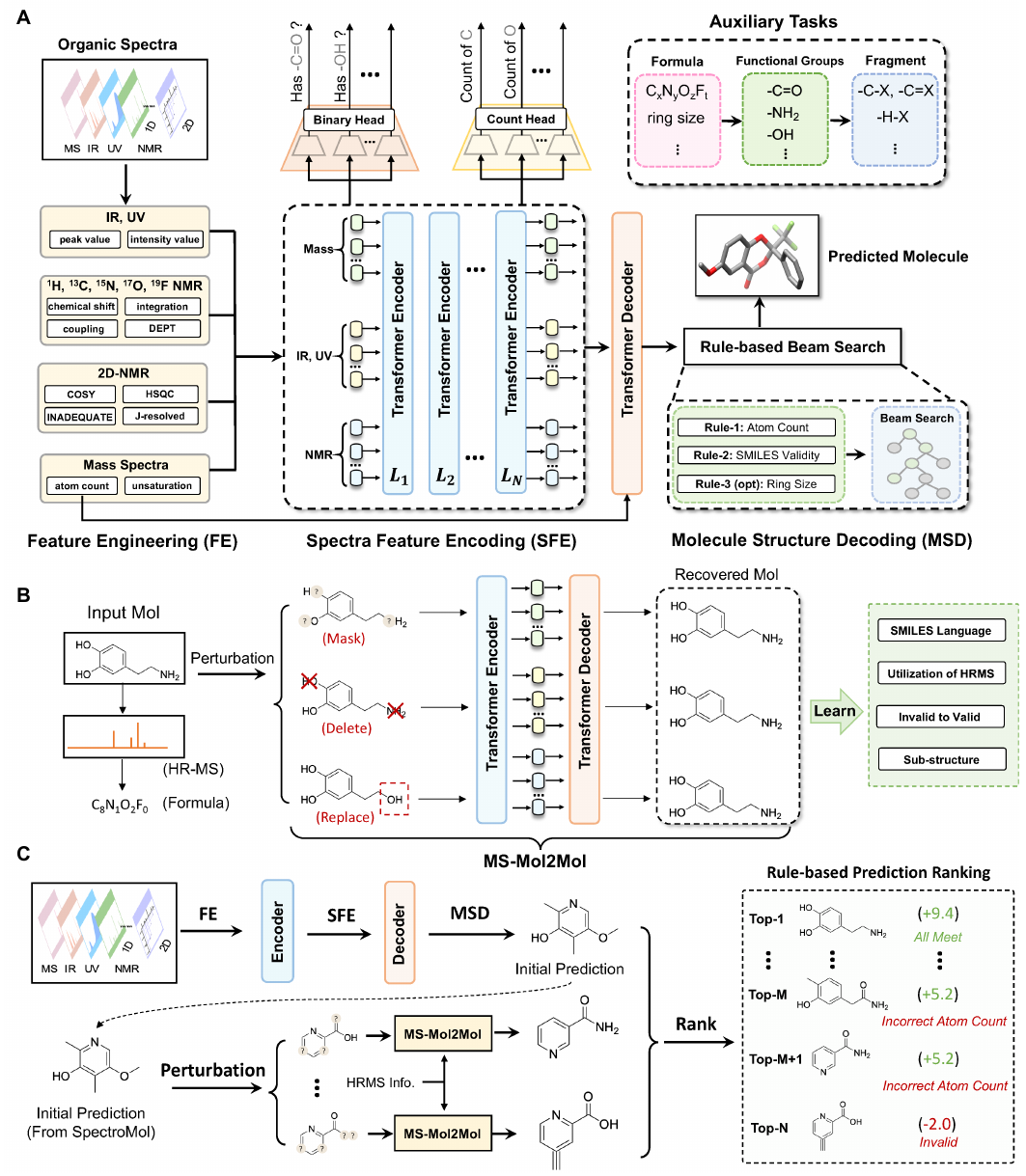}
    \caption{Framework of the propose--then--refine strategy, shown in Materials and Methods because it defines the computational protocol rather than a direct experimental result. (\textbf{A}) SpectroMol embeds IR, UV, MS, and 1D/2D NMR inputs with modality-specific encoders and fuses them in a multimodal Transformer to generate initial molecular hypotheses. (\textbf{B}) MS-Mol2Mol is pretrained by perturbation--reconstruction on 400 million molecules and conditioned on exact mass, molecular formula, and degree of unsaturation so that refinement obeys global compositional constraints. (\textbf{C}) During inference, candidate structures proposed by SpectroMol are repaired by MS-Mol2Mol and then filtered by lightweight chemical rules to remove invalid or duplicate outputs before final ranking.}
    \label{fig_2}
\end{figure}

The overall methodological design of our framework is illustrated in Figure~\ref{fig_2}. The central idea is to decouple spectral reasoning from large-scale molecular prior injection, forming a two-stage propose--then--refine paradigm that mirrors the way chemists first hypothesize a plausible scaffold and then iteratively correct it.

As shown in Figure~\ref{fig_2}a, SpectroMol performs multimodal spectral reasoning by jointly encoding IR, UV, MS, and 1D/2D NMR features within a Transformer encoder--decoder architecture. For each spectral modality $f$, the embedded representation is constructed as
\begin{equation}
\mathbf{E}_f = \mathrm{Linear}(\mathbf{X}_f) + \mathbf{P}_f + \mathbf{S}_f,
\end{equation}
where $\mathbf{X}_f$ denotes preprocessed spectral features, $\mathbf{P}_f$ positional encoding, and $\mathbf{S}_f$ spectrum-type embedding. The fused encoder memory $\mathbf{M}$ integrates complementary cross-spectral information. Conditioned on this representation, the decoder autoregressively generates a SMILES sequence $\mathbf{y}=(y_1,\dots,y_T)$ according to
\begin{equation}
p_\theta(\mathbf{y}\mid \mathbf{x})
=
\prod_{t=1}^{T}
p_\theta(y_t \mid y_{<t}, \mathbf{M}),
\end{equation}
where $\mathbf{x}$ represents the multimodal spectral input. During decoding, rule-based beam search enforces elemental budgets and syntactic validity constraints, while auxiliary prediction tasks guide the encoder toward chemically meaningful latent representations.

However, spectrum-to-structure mapping alone is fundamentally limited by spectral data scarcity and the vastness of chemical space. To address this, we introduce MS-Mol2Mol (Figure~\ref{fig_2}b), which injects large-scale molecular knowledge through perturbation--reconstruction pretraining on 400 million ZINC20 molecules. Given a corrupted SMILES $\tilde{\mathbf{y}}$ and high-resolution mass features $\mathbf{a}$, the model learns
\begin{equation}
p_\phi(\mathbf{y}\mid \tilde{\mathbf{y}}, \mathbf{a})
=
\prod_{t=1}^{T}
p_\phi(y_t \mid y_{<t}, \tilde{\mathbf{y}}, \mathbf{a}),
\end{equation}
thereby internalizing both SMILES syntax and global chemical topology. Conditioning on $\mathbf{a}$ constrains the reconstructed structure to the correct molecular formula, effectively narrowing the feasible chemical subspace and enabling correction of incomplete or imperfect hypotheses.

At inference time (Figure~\ref{fig_2}c), SpectroMol first proposes an initial candidate $\mathbf{y}^{(0)}$. Controlled perturbations generate diversified variants $\tilde{\mathbf{y}}^{(k)}$, which are refined by MS-Mol2Mol under exact-mass consistency. A lightweight rule-based filter then removes invalid or duplicate structures and ranks the refined candidates according to chemical validity and mass agreement, selecting
\begin{equation}
\hat{\mathbf{y}}
=
\arg\max_{\mathbf{y}\in\mathcal{C}}
\mathrm{Score}(\mathbf{y}),
\end{equation}
where $\mathcal{C}$ denotes the refined candidate set. 

By explicitly factorizing structure elucidation into spectral hypothesis generation followed by distribution-aware refinement, the framework leverages complementary inductive biases: cross-spectral feature integration in the first stage and large-scale molecular priors in the second. This separation is essential for scalable learning under limited spectral supervision and improves robustness under domain shift.

\subsection{Model performance under systematic ablation studies}

We first evaluate the performance of the proposed framework on the QM9-based benchmark under systematically varied spectral input conditions. Four configurations are considered: IR only, IR + UV, IR + UV + MS, and the full combination IR + UV + MS + NMR. The results consistently demonstrate that structural fidelity improves monotonically with increasing spectral diversity.

As summarized in Table~\ref{tab:spectrum_comparison}, using IR alone leads to very limited structural recovery (Top-1 accuracy 1.22\%, BLEU 0.264). Adding UV produces only marginal improvement, indicating that vibrational and electronic transitions alone are insufficient to uniquely determine molecular connectivity. The inclusion of MS significantly enhances performance (Top-1 accuracy increases to 13.4\%), reflecting the critical role of molecular formula constraints in narrowing chemical space. Strikingly, incorporating NMR information results in a dramatic performance jump, with Top-1 accuracy reaching 93.8\% and BLEU increasing to 0.970. All fingerprint- and substructure-based similarity metrics approach unity under the full spectral setting, indicating near-exact structural reconstruction.

\begin{figure}[t!]
    \centering
    \includegraphics[width=0.86\linewidth]{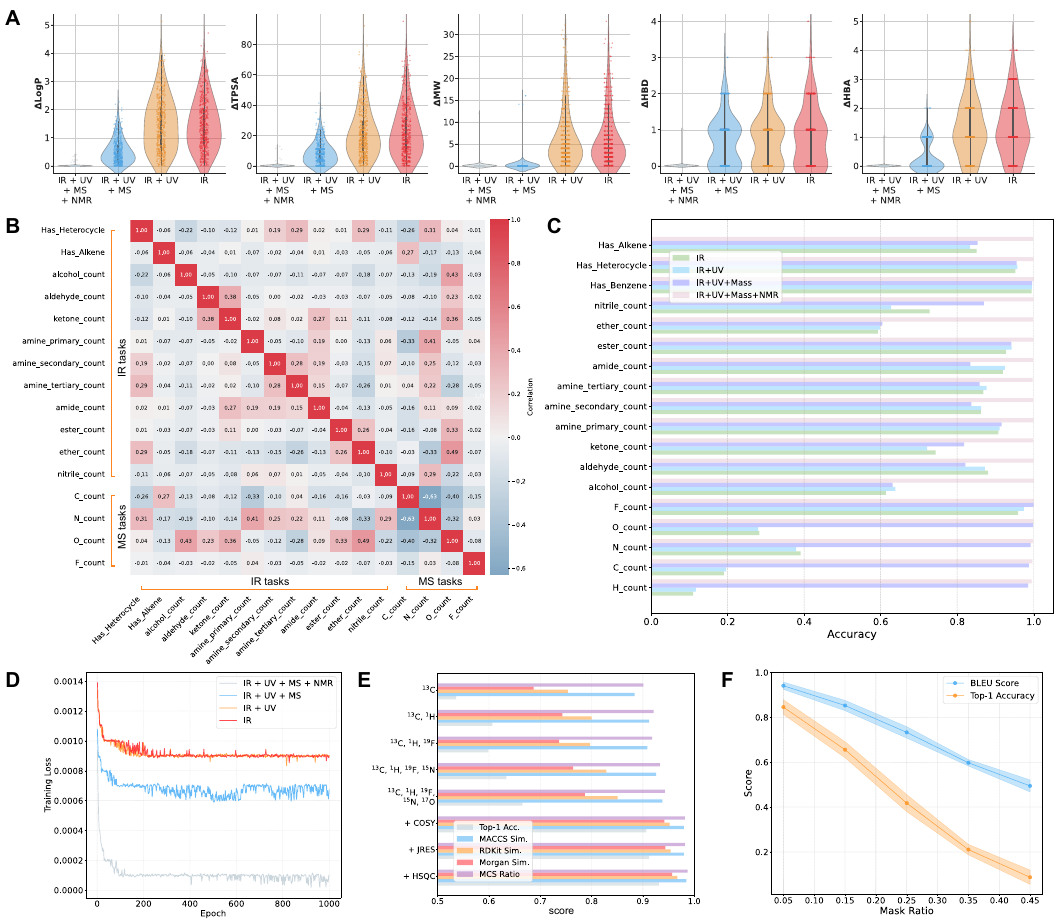}
    \caption{Systematic ablation studies on the QM9-based benchmark. (\textbf{A}) Differences in five molecular properties between generated and ground-truth molecules. (\textbf{B}) Correlations between auxiliary tasks. (\textbf{C}) Evaluation results for auxiliary prediction tasks. (\textbf{D}) Training-loss curves under different spectral combinations. (\textbf{E}) Ablation analysis of different NMR inputs. (\textbf{F}) Robustness evaluation under spectral masking.}
    \label{fig_5}
\end{figure}

The property-level deviations between predicted and ground-truth molecules further corroborate this trend (Fig.~\ref{fig_5}A). For IR-only input, the distributions of $\Delta$LogP, $\Delta$TPSA, $\Delta$MW, $\Delta$HBD, and $\Delta$HBA are broad and heavy-tailed, reflecting substantial structural ambiguity. As additional modalities are incorporated, these distributions progressively narrow. The full spectral configuration yields tightly centered violin plots with markedly reduced outliers, demonstrating that multimodal constraints effectively suppress erroneous scaffold predictions.

To understand how auxiliary tasks contribute to representation learning, we examine inter-task correlations (Fig.~\ref{fig_5}B). Moderate correlations are observed between certain IR-derived functional-group predictors and MS-based atom-count tasks, suggesting that the shared encoder captures chemically coherent latent features. The evaluation accuracy of auxiliary tasks (Fig.~\ref{fig_5}C) also improves substantially with richer spectral inputs, further supporting the role of multi-task supervision in stabilizing learning.

Training dynamics provide additional insight (Fig.~\ref{fig_5}D). Models trained with limited spectral inputs exhibit higher steady-state loss and slower convergence, whereas the full IR + UV + MS + NMR configuration achieves both faster convergence and lower final loss. This behavior indicates that complementary spectral modalities reduce ambiguity in the optimization landscape, enabling more efficient learning.

Given the dominant impact of NMR observed in Table~\ref{tab:spectrum_comparison}, we further dissect the contributions of different NMR components (Fig.~\ref{fig_5}E). Incrementally adding nuclei ($^{1}$H, $^{13}$C, $^{15}$N, $^{19}$F) and two-dimensional correlations (COSY, JRES, HSQC) consistently improve Top-1 accuracy and fingerprint similarity. The largest gains arise when 2D NMR correlations are included, highlighting the importance of connectivity-resolving information beyond one-dimensional chemical shifts.

Finally, we evaluate robustness to incomplete spectral information by randomly masking portions of the input (Fig.~\ref{fig_5}F). Both BLEU score and Top-1 accuracy decrease gradually as the masking ratio increases, but performance degradation remains modest for small perturbations ($\leq$5\%). Beyond 25--30\% masking, accuracy drops sharply, reflecting the intrinsic difficulty of structure determination under substantial information loss. Notably, even under moderate masking, the model maintains nontrivial structural similarity, suggesting that the learned cross-spectral representations exhibit resilience to partial data corruption.

Taken together, these results demonstrate that spectral modalities contribute hierarchically to structure recovery, NMR provides the decisive constraint for resolving structural degeneracy, and multi-task supervision plus multimodal fusion substantially enhance both convergence behavior and predictive robustness.

\begin{figure}[t!]
    \centering
    \includegraphics[width=0.86\linewidth]{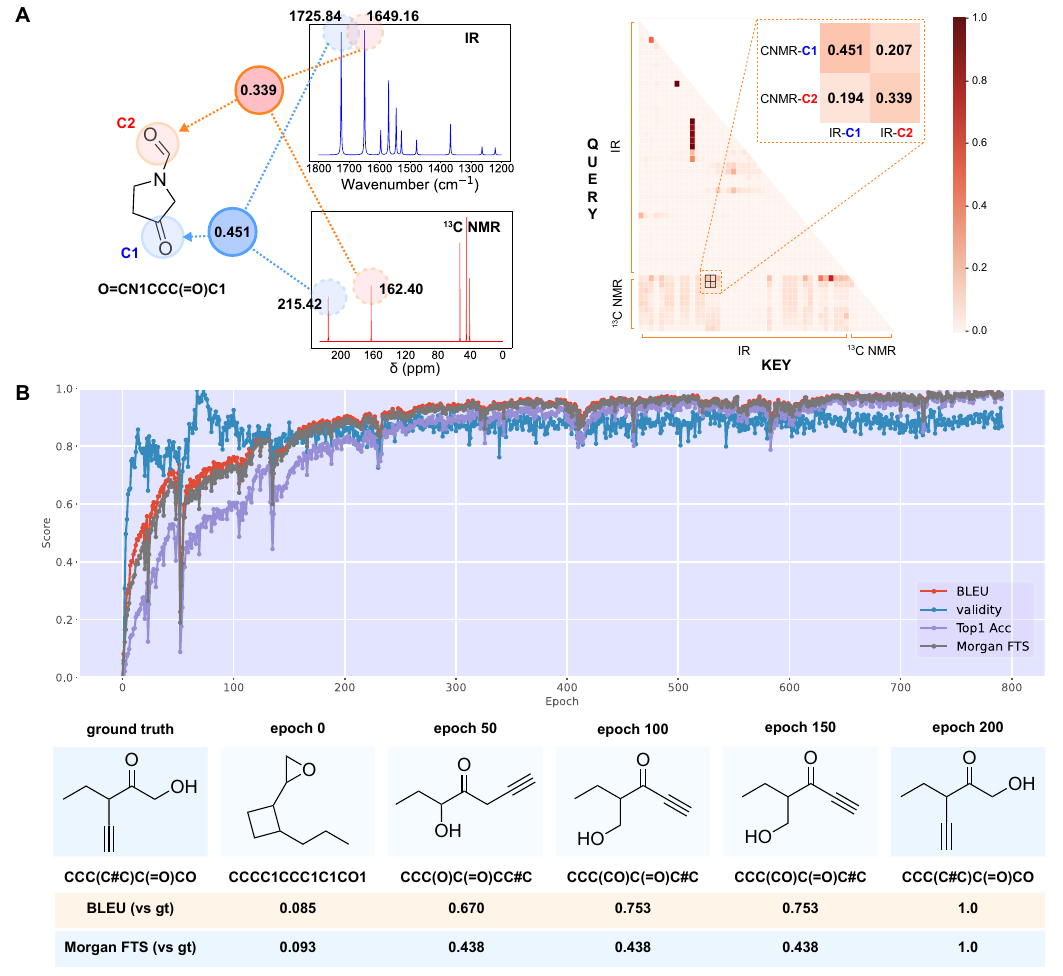}
    \caption{Interpretability analysis. (\textbf{A}) Cross-modal interaction analysis using the carbonyl group as an example, highlighting correspondence between \ce{^{13}C} NMR and IR features. (\textbf{B}) Visualization of the molecule generation process across training, with representative top predictions shown from early to late epochs.}
    \label{fig_6}
\end{figure}

\subsection{Interpretability analysis and chemical insight}

To probe whether the model learns chemically meaningful relationships rather than superficial correlations, we analyze the internal attention patterns and training dynamics shown in Fig.~\ref{fig_6}.

Figure~\ref{fig_6}A examines cross-modal interactions using the carbonyl group as a representative example. In IR spectra, carbonyl stretching vibrations typically appear in the 1600--1850~cm$^{-1}$ region, whereas in $^{13}$C NMR spectra, carbonyl carbons resonate in the 160--220~ppm range. When visualizing the cross-attention matrix between IR and $^{13}$C NMR tokens, we observe strong attention weights linking these corresponding spectral regions. For the selected molecule, the IR peaks at 1725.84 and 1649.16~cm$^{-1}$ exhibit enhanced coupling with the carbonyl carbon signals at 215.42 and 162.40~ppm. The localized high-weight blocks in the attention heatmap indicate that the model associates vibrational and chemical-shift features consistent with carbonyl functionality. This cross-spectral alignment suggests that the encoder learns chemically coherent latent representations, where complementary modalities reinforce substructure identification rather than acting independently.

Beyond static attention patterns, we examine the evolution of structure prediction during training (Fig.~\ref{fig_6}B). BLEU, Top-1 accuracy, validity, and Morgan fingerprint similarity all increase steadily, with rapid early gains followed by gradual convergence. Validity approaches unity at early epochs, whereas structural fidelity metrics improve more progressively, indicating that syntactic correctness is learned before precise connectivity. Representative molecular outputs further illustrate this progression. The initial prediction is chemically implausible, but key functional motifs emerge by epoch 50. By epoch 100, the core scaffold is largely recovered, and by epoch 200 the prediction matches the ground truth exactly. The smooth transition from invalid fragments to accurate structures reflects incremental alignment between spectral constraints and graph construction.

\begin{figure}[t!]
    \centering
    \includegraphics[width=0.8\linewidth]{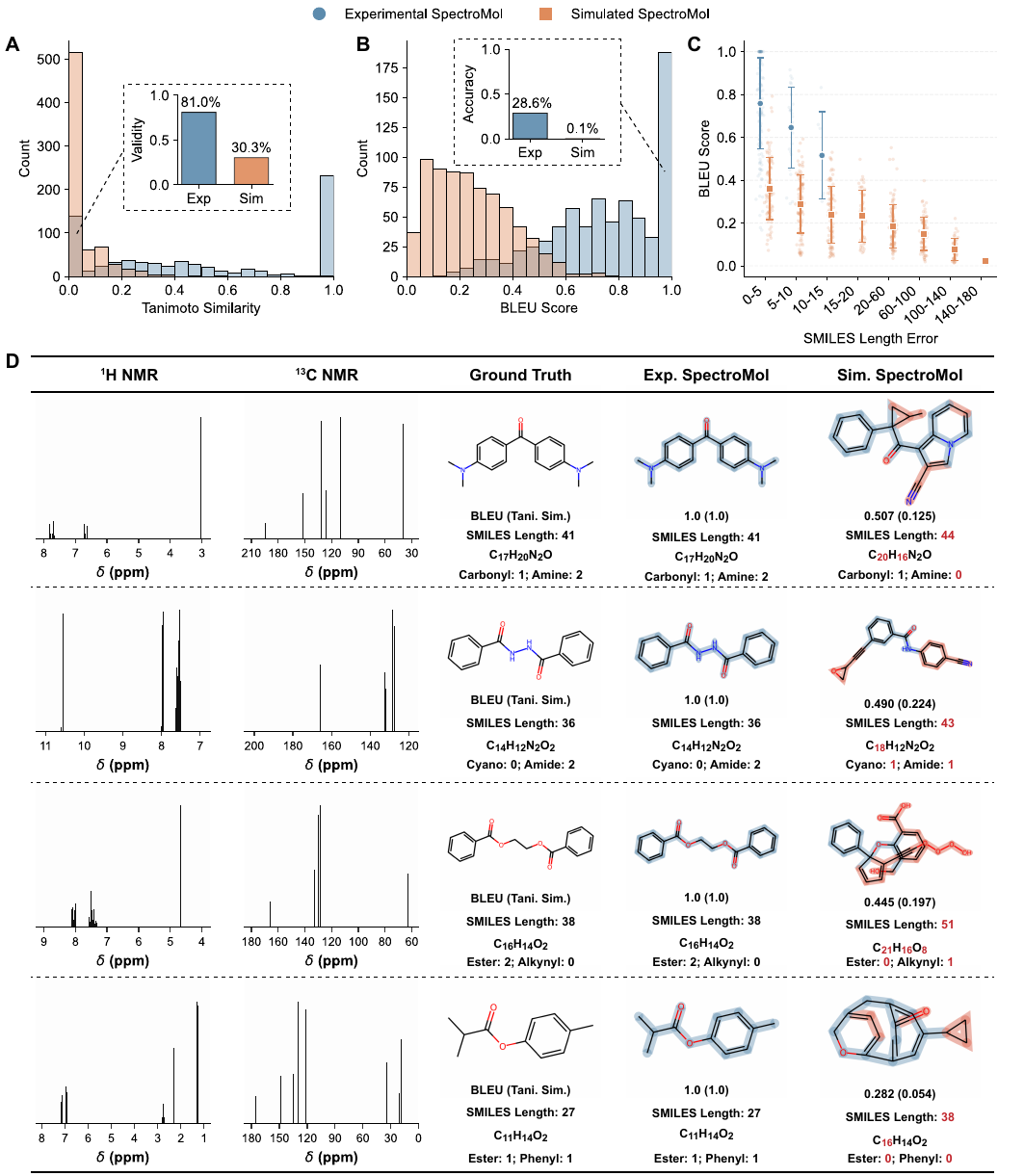}
    \caption{Adaptation from simulated to experimental spectra using only routinely accessible one-dimensional NMR and HRMS signals. (\textbf{A}) Distributions of Tanimoto similarity for structures predicted from the experimentally adapted SpectroMol model and from the simulated-spectrum baseline; inset, fraction of exact structural matches. (\textbf{B}) Corresponding BLEU-score distributions; inset, exact-reconstruction frequency. (\textbf{C}) BLEU score as a function of SMILES-length error, showing that larger deviations in inferred molecular size and complexity lead to systematically poorer sequence-level recovery. (\textbf{D}) Representative molecules comparing ground truth with predictions from the experimentally adapted model and from the simulated-spectrum baseline.}
    \label{fig_sim_exp}
\end{figure}

\subsection{Adapting from simulated to experimental spectra}

A central practical question is whether a model trained mainly on inexpensive simulated spectra can be transferred to real laboratory measurements, where line broadening, peak overlap, missing signals, and instrument-dependent distortions are unavoidable. Because curated experimental multimodal datasets remain scarce---especially for two-dimensional NMR---we restricted this analysis to the spectral information most commonly obtainable in practice: $^1$H NMR, $^{13}$C NMR, and high-resolution mass spectrometry. We assembled 10,000 experimental training examples from the AIST Spectral Database for Organic Compounds (SDBS)~\cite{exp_aist}, initialized the model from the simulated-spectrum checkpoint, and continued training for 100 epochs under the same optimization settings. To better bridge the gap between QM9-scale molecules and larger literature compounds, the simulated pretraining stage was aligned with one-dimensional spectra from the multimodal NIPS dataset of Alberts et al.~\cite{alberts2024unraveling}. This setting therefore asks a methodologically important question: can broad pretraining on cheap computed spectra provide a transferable prior that is later calibrated with only a modest amount of experimental data?

The answer is affirmative, but the transfer is chemically nontrivial. As shown in Fig.~\ref{fig_sim_exp}A and B, experimental fine-tuning shifts the prediction distribution toward substantially higher structural similarity, whereas the simulated-spectrum baseline remains concentrated in low-similarity and low-BLEU regimes. Exact Tanimoto recovery reaches 30.3\% for the experimentally adapted model, compared with only 0.1\% for the simulated-spectrum baseline, while the frequency of exact BLEU reconstruction rises from 0.1\% to 28.6\%. Figure~\ref{fig_sim_exp}C further shows that BLEU decays progressively as SMILES-length error grows. This trend indicates that experimental failures are often accompanied by systematic mistakes in inferred molecular size, ring count, or overall topological complexity, rather than by only a few misplaced tokens. Chemically, this is consistent with a model that has already learned local environments from simulated spectra but is not yet fully calibrated to real instrumental line shapes and peak distributions.

\begin{figure}[t!]
    \centering
    \includegraphics[width=0.96\linewidth]{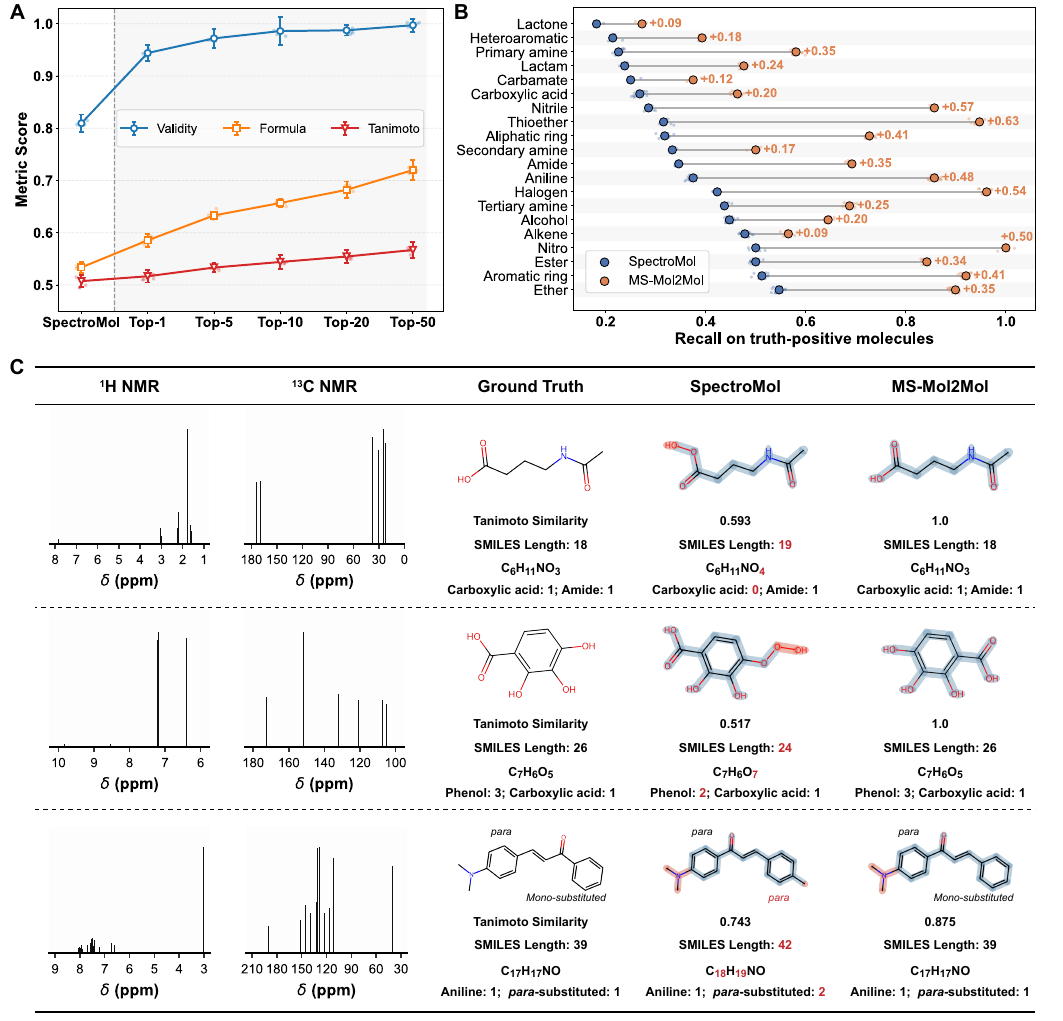}
    \caption{MS-Mol2Mol repair of experimental-spectrum predictions. (\textbf{A}) Top-$k$ evaluation of validity, molecular-formula correctness, and Tanimoto similarity before and after refinement, showing that mass-guided repair drives validity to near saturation and markedly improves formula consistency. (\textbf{B}) Functional-group recall on true-positive molecules for representative motifs, comparing the initial SpectroMol hypotheses with the refined MS-Mol2Mol outputs. (\textbf{C}) Representative experimental cases in which MS-Mol2Mol corrects composition errors and local topology while preserving the structurally relevant information already captured by SpectroMol.}
    \label{fig_exp_refine}
\end{figure}

The molecule-level examples in Fig.~\ref{fig_sim_exp}D make this point more concrete. In the first example, the target is a carbonyl-containing aromatic system bearing two amine substituents. The experimentally adapted model recovers the exact structure, preserving both the oxidation pattern and the nitrogen count. By contrast, the simulated-spectrum baseline drifts toward a larger heteroaromatic scaffold with the wrong carbonyl context and a different heteroatom arrangement. This difference is chemically meaningful because carbonyl-containing aromatic systems can produce broadly similar downfield $^{13}$C signals, yet the joint pattern of aromatic proton environments and exact mass should distinguish a compact diaryl carbonyl motif from an overextended fused alternative. In the second example, the experimentally adapted model again reproduces the exact target, whereas the simulated-spectrum baseline introduces a cyano-containing alternative and changes the amide balance. Here the essential failure is heteroatom misallocation: the model recognizes that the molecule is nitrogen- and oxygen-rich, but without experimental calibration it places one nitrogen into a nitrile-like environment rather than preserving the correct amide-rich arrangement. The third example is equally informative. The target contains two ester functionalities but no alkynyl unit. The experimentally adapted model returns the exact ester-containing constitution, while the simulated-spectrum baseline invents an alkynyl motif and simultaneously loses the proper oxygenation pattern. This kind of error reflects a chemically familiar ambiguity: unsaturation inferred from sparse one-dimensional information can be distributed incorrectly between carbonyl-derived unsaturation and carbon--carbon multiple bonds. The fourth example shows a similar trend in a smaller aryl ester. The adapted model correctly identifies the target framework, whereas the simulated baseline erases both the ester count and the phenyl assignment. Across all four cases, experimental adaptation improves three particularly important chemical judgments: the allocation of heteroatoms, the assignment of oxidation state, and the placement of unsaturation within aromatic frameworks.

\subsection{MS-guided refinement of experimental-spectrum predictions}

We next asked whether the remaining experimental errors could be corrected by a second-stage model that explicitly enforces mass consistency. Figure~\ref{fig_exp_refine}A shows that MS-Mol2Mol is highly effective in exactly the aspects that matter most to practicing chemists. Across the top-$k$ candidate lists, validity is driven essentially to saturation, no-hydrogen molecular-formula correctness rises sharply, and Tanimoto similarity also improves. By contrast, BLEU is not the dominant story after refinement. This behavior is chemically sensible. Once refinement begins to enforce exact mass, degree of unsaturation, and permissible elemental composition, the selected structure need not maximize token overlap with the ground-truth SMILES; it only needs to move toward the correct scaffold neighborhood while satisfying the harder constraints imposed by HRMS. In real structure elucidation, a constitutionally valid and compositionally correct candidate is far more informative than a textually similar but chemically impossible string.

The functional-group analysis in Fig.~\ref{fig_exp_refine}B clarifies why the refinement stage is useful. The largest gains are observed for groups such as carboxylic acids, amides, phenols, esters, lactones, and aromatic substitution classes. These are precisely the motifs for which local NMR evidence alone can leave chemically plausible ambiguity. For instance, $^1$H and $^{13}$C NMR may strongly indicate an oxygenated aromatic framework, but they do not always uniquely determine whether a given oxygen belongs to a phenol, ester, lactone, or carboxylic acid, especially when peak overlap and experimental noise are present. HRMS-derived formula constraints become decisive in these settings because one additional oxygen or one reallocated nitrogen can completely change the functional-group identity even if the surrounding scaffold remains broadly similar. MS-Mol2Mol exploits this global bookkeeping role of mass information and therefore preferentially improves motifs whose correct recovery depends on accurate heteroatom counting and valence balance.

The representative cases in Fig.~\ref{fig_exp_refine}C illustrate these effects in chemically intuitive terms. In the first example, the target is an oxygen-rich aromatic acid containing three phenolic oxygens and one carboxylic acid. SpectroMol already places the molecule in the correct family, but it underestimates oxygenation and returns a structure with only two phenolic oxygens. MS-Mol2Mol restores the full oxidation pattern and recovers the exact target. In the second example, the target contains one carboxylic acid and one amide. Before refinement, SpectroMol proposes an overoxidized analogue with the wrong oxygen count, effectively shifting the functional-group balance even though the broad scaffold remains related. After refinement, the acid/amide partition and the overall composition are both corrected. The third example is particularly instructive because refinement does not find the exact target yet still produces a chemically better hypothesis. The ground truth is an aniline-containing aromatic framework with a defined para-substitution relationship. SpectroMol captures the rough scaffold but overpredicts para substitution and perturbs the aromatic arrangement. MS-Mol2Mol moves the prediction closer to the true substitution logic and improves Tanimoto similarity from 0.743 to 0.875, even though exact reconstruction is not achieved. This behavior is important: refinement succeeds not only when it finds the exact molecule, but also when it repairs heteroatom bookkeeping and substitution logic enough to produce a more credible structural hypothesis.

Overall, these results support a chemically intuitive division of labor. SpectroMol serves as the spectral interpreter that extracts scaffold-level and motif-level hypotheses from NMR and HRMS, whereas MS-Mol2Mol serves as the composition-aware repair engine that enforces exact-mass consistency and improves formula correctness. This combination is particularly attractive for practical deployment because computed spectra are cheap and scalable for pretraining, experimental spectra are expensive but highly informative for calibration, and mass-guided refinement provides a final safeguard against chemically inadmissible predictions.

\section{Conclusion}
\label{conclusion}

We have presented a multimodal propose--then--refine framework for automated organic structure elucidation that separates spectral interpretation from composition-aware molecular correction. Within this design, SpectroMol converts heterogeneous IR, UV, HRMS, and 1D/2D NMR measurements into chemically plausible structural hypotheses, whereas MS-Mol2Mol uses exact-mass constraints together with a large-scale generative prior to repair compositional inconsistencies and sharpen constitutional assignments. The resulting workflow moves beyond single-stage spectrum-to-SMILES translation toward a form of constrained chemical reasoning in which local spectroscopic evidence and global molecular plausibility are explicitly coordinated.

Several broader implications emerge from this study. First, the results show that multimodal gains arise from complementarity rather than simple data accumulation: NMR resolves local connectivity, HRMS enforces compositional admissibility, and their combination markedly reduces structural degeneracy. Second, simulated spectra can provide a scalable pretraining substrate for inverse chemical modeling, while limited experimental data serve to calibrate the model to laboratory reality. Third, refinement under exact-mass constraints improves not only formal validity but also the chemical usefulness of the final hypotheses, which is essential for practical deployment in synthesis, analysis, and autonomous experimentation. Although larger experimental datasets, richer two-dimensional measurements, and mixture-level settings remain important future directions, the present work establishes a realistic route toward data-driven structure elucidation that is scalable, chemically informed, and compatible with real-world spectroscopic workflows.

\section{AUTHOR INFORMATION}

\subsection{Corresponding Authors}

\textbf{Li Yuan} -- 
School of AI for Science, Peking University Shenzhen Graduate School, Shenzhen 518055, China; 
School of Electronic and Computer Engineering, Peking University Shenzhen Graduate School, Shenzhen 518055, China; 
Peng Cheng Laboratory, Shenzhen 518000, China.  
Email: yuanli-ece@pku.edu.cn

\vspace{6pt}

\textbf{Bartosz A. Grzybowski} -- 
Institute of Organic Chemistry, Polish Academy of Sciences, ul. Kasprzaka 44/52, Warsaw 01-224, Poland; 
IBS Center for Algorithmic and Robotized Synthesis (CARS), 50 UNIST-gil, Eonyang-eup, Ulju-gun, Ulsan 689-798, South Korea; 
Department of Chemistry, UNIST, 50 UNIST-gil, Eonyang-eup, Ulju-gun, Ulsan 689-798, South Korea.  
ORCID: 0000-0001-6613-4261
Email: nanogrzybowski@gmail.com

\vspace{6pt}

\textbf{Fanyang Mo} -- 
School of AI for Science, Peking University Shenzhen Graduate School, Shenzhen 518055, China; 
AI for Science (AI4S)-Preferred Program, Peking University Shenzhen Graduate School, Shenzhen 518055, China; 
State Key Laboratory of Advanced Waterproof Materials, School of Materials Science and Engineering, Peking University, Beijing 100871, China; 
School of Advanced Materials, Peking University Shenzhen Graduate School, Shenzhen 518055, China; 
Guangdong Provincial Key Laboratory of Nano-Micro Materials Research, Peking University Shenzhen Graduate School, Shenzhen 518055, China.  
ORCID: 0000-0002-4140-3020  
Email: fmo@pku.edu.cn

\subsection{Authors}

\textbf{Chengchun Liu} -- 
School of AI for Science, Peking University Shenzhen Graduate School, Shenzhen 518055, China; 
School of Materials Science and Engineering, Peking University, Beijing 100871, China.  
ORCID: 0009-0002-5550-4145

\vspace{6pt}

\textbf{Zhiyuan Yan} -- 
School of AI for Science, Peking University Shenzhen Graduate School, Shenzhen 518055, China; 
School of Electronic and Computer Engineering, Peking University Shenzhen Graduate School, Shenzhen 518055, China.

\vspace{6pt}

\textbf{Hao Li} -- 
School of AI for Science, Peking University Shenzhen Graduate School, Shenzhen 518055, China; 
School of Electronic and Computer Engineering, Peking University Shenzhen Graduate School, Shenzhen 518055, China.

\vspace{6pt}

\textbf{Boxuan Zhao} -- 
School of AI for Science, Peking University Shenzhen Graduate School, Shenzhen 518055, China; 
School of Materials Science and Engineering, Peking University, Beijing 100871, China.

\vspace{6pt}

\textbf{Yonghong Tian} -- 
School of AI for Science, Peking University Shenzhen Graduate School, Shenzhen 518055, China; 
School of Electronic and Computer Engineering, Peking University Shenzhen Graduate School, Shenzhen 518055, China; 
Peng Cheng Laboratory, Shenzhen 518000, China.

\subsection{Author Contributions}

C.L. and Z.Y. contributed equally to this work.  
C.L. and Z.Y. conceived the methodological frameworks.  
C.L. and F.M. performed large-scale computational data generation, including DFT-based spectral calculations and dataset construction.  
C.L., Z.Y., and F.M. conducted the chemical interpretability analysis and theoretical validation of the spectral–structure relationships.  
H.L., B.Z., and Y.T. provided critical feedback and contributed to methodological refinement.  
L.Y., B.A.G., and F.M. supervised the project and provided strategic guidance on methodology development and chemical applications.

\subsection{Notes}

The authors declare no competing financial interest.

\section{ACKNOWLEDGMENTS}

We thank Peking University Shenzhen Graduate School and Shenzhen Government for the start-up funding support. We thank the High-Performance Computing Platform of Peking University for machine learning model training.

\bibliography{example-refs}

\end{document}